\pdfoutput=1
\documentclass{article}
\usepackage{amsmath}
\usepackage{amssymb}
\usepackage{subcaption}
\usepackage{hyperref}
\usepackage{cleveref}
\usepackage{lineno}
\usepackage{color}
\usepackage{graphicx}
\usepackage{fullpage}
\usepackage{url}
\usepackage[utf8]{inputenc}
\usepackage[english]{babel} % English language/hyphenation
\usepackage{bm} 
\usepackage[bottom]{footmisc}
\usepackage{xargs}

\newcommandx{\citep}[3][1=,2=]{#1\cite{#3}#2}
\newcommand{\lam}{\lambda}

\newcommand{\PD}[2]{\frac{\partial #1}{\partial #2}}

\newcommand{\algn}[1]{\begin{linenomath}\begin{align} #1 \end{align}\end{linenomath}}
\newcommand{\algnn}[1]{\begin{linenomath}\begin{align*} #1 \end{align*}\end{linenomath}}
\newcommand{\B}[1]{\mathbf{#1}}

\newcommand{\arr}{\rightarrow}

\newcommand{\half}{\tfrac{1}{2}}
\newcommand{\bs}[1]{\boldsymbol{#1}}
\newcommand{\tl}[1]{\tilde{#1}}
\newcommand{\EXP}[1]{\left\langle #1 \right\rangle}

\newcommand{\lrb}[1]{\left(#1\right)}
\newcommand{\lrsq}[1]{\left[#1\right]}
\newcommand{\lrc}[1]{\left\{#1\right\}}
\newcommand{\al}{\alpha}
\newcommand{\Half}{\frac{1}{2}}

\newcommand{\eqsys}[1]{\begin{subequations}\algn{#1}\end{subequations}}

\newcommand{\eps}{\epsilon}
\newcommand{\be}{\beta}
\newcommand{\abs}[1]{\left| #1 \right|}

\newcommand{\rmd}{ {\color{red} {\rm d}}}
\renewcommand{\L}{\mathcal{L}}
\newcommand{\dD}[2]{\frac{\delta #1}{\delta #2}}

\DeclareMathOperator{\curl}{curl}

\title{Variational Stochastic Parameterisations and their Applications to Primitive Equation Models}
\author{Ruiao Hu and Stuart Patching
\footnote{Corresponding author, email: s.patching17@imperial.ac.uk}\\
	ruiao.hu15@imperial.ac.uk, s.patching17@imperial.ac.uk\\
	Department of Mathematics, Imperial College London \\ SW7 2AZ, London, UK}

\begin{document}

\iffalse
%
% RECOMMENDED %%%%%%%%%%%%%%%%%%%%%%%%%%%%%%%%%%%%%%%%%%%%%%%%%%%
%

\def\UrlFont{\rmfamily}
\input{Preamble}

\begin{document}

\mainmatter              % start of a contribution
	%
\title{Variational Stochastic Parameterisations and their Applications to Primitive Equation Models}
	%
\titlerunning{Variational Stochastic Parameterisations}  % abbreviated title (for running head)
\author{Ruiao Hu\inst{1} \and Stuart Patching\inst{2}}
%
\authorrunning{} % abbreviated author list (for running head)
\tocauthor{Ruiao Hu and Stuart Patching}
%
\institute{Department of Mathematics, Imperial College London \\ SW7 2AZ, London, UK \\
	\email{ruiao.hu15@imperial.ac.uk}
	\and 
	Department of Mathematics, Imperial College London \\ SW7 2AZ, London, UK \\
	\email{s.patching17@imperial.ac.uk}
	\footnote{Corresponding author, email: s.patching17@imperial.ac.uk}
	}

\fi

\maketitle              % typeset the title of the contribution

\begin{abstract}
We present a numerical investigation into the stochastic parameterisations of the Primitive Equations (PE) using the Stochastic Advection by Lie Transport (SALT) and Stochastic Forcing by Lie Transport (SFLT) frameworks. These frameworks were chosen due to their structure-preserving introduction of stochasticity, which decomposes the transport velocity and fluid momentum into their drift and stochastic parts, respectively. In this paper, we develop a new calibration methodology to implement the momentum decomposition of SFLT and compare with the Lagrangian path methodology implemented for SALT. The resulting stochastic Primitive Equations are then integrated numerically using a modification of the FESOM2 code.
For certain choices of the stochastic parameters, we show that SALT causes an increase in the eddy kinetic energy field and an improvement in the spatial spectrum. SFLT also shows improvements in these areas, though to a lesser extent. SALT does, however, have the drawback of an excessive downwards diffusion of temperature. 

%\keywords{Primitive equations, Geometric mechanics, FESOM2,\\ Stochastic Parameterisation}
\end{abstract}

\section{Introduction}
Uncertainty can be present in ocean models due to a number of factors including, but not limited to: small-scale processes not resolved by the grid; observation error; model error; numerical error and unrealistic viscosities imposed to ensure numerical stability. Several stochastic parameterisation techniques \citep{mana2014toward,berloff_2005,memin2014,holm2015,HolmHu2021} have been proposed recently as ways of representing uncertainty in ocean models. Because these parameterisations are probabilistic, it is possible to generate ensemble forecasts \citep{cotter2019numerically,cotter2018modelling,cotter2020data,uchida2021ensemble} with associated means and variances, which can then be applied to data assimilation.  %{\color{red} These parameterisation are \emph{probabilistic} where the resulting models can predict both the expected and the variance of the state variables. These stochastic models allow us to generate ensemble forecasts \citep{cotter2019numerically,cotter2018modelling,cotter2020data,uchida2021ensemble} which can then be applied to data assimilation.} 
This work will focus on two frameworks which introduce stochasticity in a way that preserves certain fundamental and desirable properties of fluid flows. These frameworks are: Stochastic Advection by Lie Transport (SALT) \citep{holm2015} and Stochastic Forcing by Lie Transport (SFLT) \citep{HolmHu2021}. Both SALT and SFLT are derived from variational principles, from which we may observe the geometric structure of the fluid equations and the conservation laws which are inherited.\par
The key assumption of SALT is the decomposition of transport velocity into a slow mean part and a fast, rapidly fluctuating part around the mean. In the limit of high fluctuation frequency, one can use homogenisation theory to transform the rapidly-fluctuating component to a sum of stochastic vector fields \citep{CGH17}. Thus, the modification from the deterministic flow is the addition of stochastic vector fields to the transport velocity. This stochastic modification has been shown \citep{holm2015} to preserve the Kelvin circulation theorem and the advection equation for potential vorticity. In the case where buoyancy obeys an advection relation, the potential vorticity is conserved along particle paths. However, SALT violates energy conservation %of deterministic Hamiltonian systems
since stochastic Hamiltonians are introduced into the variational principle. The application of the SALT in quasi-geostrophic (QG) models and the 2D Euler equations has been investigated before in \citep{cotter2018modelling,cotter2019numerically,cotter2020data}. However, these models are too simplistic to be used in operational ocean simulations, and the majority of ocean codes (e.g. MOM5 \citep{GRIFFIES2000123}, ICON \citep{KORN2017525}, MITgcm \citep{marshall_adcroft_hill_perelman_heisey_1997}, FESOM2 \citep{danilov_sidorenko_wang_jung_2016}) solve the Primitive Equations (PE). For this reason, if SALT is to be employed for use in practical applications, it must be adapted for use in PE. This introduces additional features to the model as compared the QG or 2D Euler: in PE there are advected quantities such as temperature and salinity, which in the SALT framework are advected by the stochastic velocity. There is, moreover, a subtlety in the pressure arising from the imposition of a semi-martingale Lagrange multiplier in the incompressibility condition of the variational principle \citep{street2021}.
\par
An alternative stochasatic parameterisation is the more recent SFLT framework \citep{HolmHu2021}. Derived via a Lagrange-d'Alembert principle, SFLT allows the addition of arbitrary stochastic forcings to the evolution equations of the momentum and of the advected quantities. This modification differs from SALT, as stochasticity is added in the variational principle \emph{after} taking variations of the Hamiltonian for the deterministic system . %Thus, the Hamiltonian, which is equal to the total energy, remains that of its deterministic counterpart, as given in \cref{eq:hamiltonian}. 
By considering the Lie-Poisson bracket of the system, we choose the forcing to be of a particular form that preserves, on every realisation of the noise, the original (deterministic) Hamiltonian. For PE, the Hamiltonian is given in \cref{eq:hamiltonian}.
However, the addition of energy preserving forces will modify the Kelvin circulation theorem. In the current work, we will consider the case where the stochastic forcing is in the energy conserving form and applied to the momentum equation.
%It is, in theory, possible to introduce a stochastic energy-preserving forcing on the advected quantities; we will not, however, investigate this in detail here.
As in the SALT case, stochastic pressure terms will appear in the momentum equation due to the imposition of semi-martingale Lagrange multiplier in the incompressibility constraint. Prior to the present work, SFLT has not been implemented into numerical models.\par 
The rest of the paper is structured as follows. In \cref{sec:VPs}, we derive PE with both SALT and SFLT from a variational principle and we show the conservation properties from the resulting equations. In \cref{sec:Xis}, we consider calibration procedures to calculate the stochastic parameters of SALT and SFLT. In particular, we use the Lagrangian paths method of \citep{cotter2018modelling} but also consider a simpler technique, that of Eulerian differences, which we propose is more appropriate for use in SFLT. In \cref{sec:results}, we present numerical results of applying SALT and SFLT to FESOM2 \citep{danilov_sidorenko_wang_jung_2016} (see \cref{sec:Numerics}), demonstrating the different effects of these stochastic frameworks and the sensitivity to the choice of parameters.

% We demonstrate how to derive PE with SALT and SFLT from a variational principle and we show the conservation properties of the resulting equations. We propose methods for calculating the parameters of SALT and SFLT. In particular, we use the Lagrangian paths method of \citep{cotter2018modelling} but also consider a simpler technique, that of Eulerian differences, which we propose is more appropriate for use in SFLT. We outline how SALT and SFLT may be introduced into the ocean model FESOM2 \citep{danilov_sidorenko_wang_jung_2016}. Some preliminary numerical results are presented, demonstrating the different effects of SALT and SFLT, and the sensitivity to the choice of parameters. 

\section{Stochastic Primitive Equations}
\label{sec:VPs}
\subsection{Variational Principles for Stochastic Primitive Equations}
%\begin{framed} {\color{red} RH: Should we write use the red $\rmd$ to mean a stochastic time derivative to distinguish from the $\rmd t$ and $\rmd W$ semi-martingales?}\end{framed}
Variational principles may be used to derive systems of fluid equations \citep{HMR1998,holm2009geometric} which obey conservation laws such as the Kelvin circulation theorem. %In order to derive a system of equations from a variational principle, we need to know what the appropriate Lagrangian is. It turns out \citep{holm2009geometric} that for the Primitive Equations the correct choice is:
To derive the Primitive Equations from a variational principle, the appropriate Lagrangian is \citep{holm2009geometric}:
\algn{
    l(\B{u}, D, T,S) =  \int \lrb{\Half \abs{\B{u}}^2 + \B{u}\cdot\B{R} - V(T,S,z)} D d^3 x \,,
}
where $\B{u} = (u,v)$ is the horizontal velocity vector field, $\B{R}$ is the Coriolis potential, which satisfies $\curl \B{R} = f(y) \hat{\B{z}}$ with $f(y)=2\Omega \cos y$ and $\Omega = 2\pi/\rm{day}$ is the rotational frequency of the earth. $T$ and $S$ are the temperature and salinity respectively; these are tracers advected by the fluid. $D$ is the Jacobian of the flow map $g_t$ that maps a fluid particle at initial position $\B{x}_0$ to its position $\B{x}_t=g_t\B{x}_0$ at time $t$. $V$ is the potential energy, which has explicit dependence on $T$ and $S$, as well as on the vertical coordinate $z$. The three-dimensional velocity shall be denoted $\B{v} = (\B{u},w)$. \par 
In order to obtain the correct hydrostatic balance condition the potential energy should obey $\PD{V}{z}(T,S,z)= g(1+b)$ where the partial derivative is taken with respect to $z$ at constant $T,S$. $b$ is the buoyancy, given by the equation of state $b=b(T,S,z)$.\par
It is convenient here to use the Clebsch version of the variational principle \citep{Cotter_holm_2009} in Hamiltonian form. The Hamiltonian is given by Legendre transformation as $h(\B{m}^h, D, T, S):= \EXP{ \B{u},\dD{l}{\B{u}}} - l(\B{u}, D, T, S)$ where $\B{m}^h:=\dD{l}{\B{u}} = D\lrb{\B{u} + \B{R}}$ is the horizontal momentum.
We have also defined the inner product $\EXP{p,q} = \int p\cdot q d^3x$. We shall use the same angle-bracket notation for all such pairings, when $p$ and $q$ are dual variables, e.g. vector field and 1-form density; or a scalar and a density.
The Hamiltonian can be written explicitly as:
\algn{\label{eq:hamiltonian}
    h(\B{m}^h, D,T,S) = \int \lrb{ \Half\abs{\frac{\B{m}^h}{D} - \B{R}}^2 + V(T,S,z)}D d^3 x \, .
}
% In the Clebsch variational principle, we define the (3-dimensional) transport velocity as a stochastic process, $\rmd \bs{\chi}$. This is done by including in the action at least three transport equations of the form $\lrb{\rmd  + \L_{\rmd \bs{\chi}}}a = 0$, where $a \in \{D,T,S\}$; these equations are imposed by Lagrange-multiplier constraints, see \citep{Seliger1968}.
In the Clebsch variational principle when SALT or SFLT are present, the (3-dimensional) transport velocity $\rmd \bs{\chi}$ is defined to be a stochastic process. The form of $\rmd \bs{\chi}$ is defined using Lagrange-multiplier constraints to impose the transport equations $\lrb{\rmd  + \L_{\rmd \bs{\chi}}}a = 0$, where $a \in \{D,T,S\}$, \citep[see ][]{Seliger1968}.
Here we remark that for clarity, we denote by an italic $d$ the spatial differential and a straight red $\rmd$ for the stochastic time-increment.  $\L_{\rmd \bs{\chi}}$ denotes the Lie derivative, which is a differential operator with a form that depends on the object on which it acts. We remark here that there is a slight abuse of notation and we shall write $D$ as a short-hand for $D d^3x$ so that this is a density 3-form and the Lie derivative is given by $\L_{\rmd \bs{\chi}}D = \nabla\cdot\lrb{\rmd \bs{\chi}\, D}$. $T$ and $S$ are scalars, so we have $\L_{\rmd \bs{\chi}}T := \rmd \bs{\chi} \cdot \nabla T$ and similarly for $S$. 
In order to obtain the incompressibility of the transport velocity $\rmd\bs{\chi}$, we include an additional constraint to set $D=1$ where the Lagrange multiplier will be interpreted as the pressure. Since the Hamiltonian $h$ only depends on the horizontal momentum $\B{m}^h$, we need to include an extra constraint so that the vertical component of the momentum is set to zero; this will give us hydrostatic balance. \\
\par 
The defining feature of SALT is that the transport velocity is the sum of the drift velocity and a number of stochastic corrections to the drift:
\algn{\label{eq:chi_def}
\rmd \bs{\chi}(\B{x},t) := \B{v}(\B{x},t)\rmd t + \sum_i \bs{\xi}_i(\B{x},t) \circ \rmd W_t^i \,,
}
where $\bs{\xi}_i(\B{x}, t)$ are arbitrary vector fields. We remark here that \cref{eq:chi_def} is a stochastic process at fixed Eulerian points $\B{x}$ and we do not solve for this process explicitly. $\rmd\bs{\chi}$ is distinct from the particle trajectories $\B{x}_t$, which evolve in time according to $\rmd \B{x}_t = \B{v}(\B{x}_t,t)\rmd t + \sum_i \bs{\xi}_i(\B{x}_t,t) \circ \rmd W_t^i$ and will be used during calibration procedures in \cref{sec:Xis}. We can impose the form of the transport velocity specified in \cref{eq:chi_def} by including in the action some additional stochastic Hamiltonians $\sum_i h_i(\B{m}^h)\circ \rmd W_t^i$ where the horizontal component of the parameters is given by $\bs{\xi}_i^h(\B{x},t) = \dD{h_i}{\B{m}^h}$. The three-dimensional momentum is denoted $\B{m}=(\B{m}^h, m_3)$. We note that in principle $\bs{\xi}_i$ may depend on time; however, we shall henceforth assume for simplicity that $\bs{\xi}_i = \bs{\xi}_{i}(\B{x})$ is a function of space only. When $h_i$ are independent of $\B{m}^h$, we have the relation $\rmd \bs{\chi}(\B{x},t) := \B{v}(\B{x},t)\rmd t$, so that $\rmd\bs{\chi}$ reduces to the original deterministic transport. \\ 
\par
SFLT is included \citep{HolmHu2021} via a Lagrange-d'Alembert term $\EXP{\delta \rmd\bs{\chi}, \B{F}}$ added to the variation of the action $\delta S$. Since this is added after variations of the action are taken, the forcing $\B{F}$ can in principle be arbitrary. % We can also add arbitrary forcings to the advection equations for $D,T,S$, though we do not consider these here. 
Overall, the variational principle takes the following form:
\algn{
\begin{split} \label{eq:action_SALT_SFLT}
    0 = \delta \mathcal{S} = \delta &\left. \int \EXP{\rmd \bs{\chi},\B{m}} - h(\B{m}^{(h)},D,T,S)\rmd t - \EXP{\rmd \zeta,m_3} - \EXP{\rmd P, D-1 } \right.\\
    &\quad\left.+\EXP{\al, \lrb{\rmd + \mathcal{L}_{\rmd \bs{\chi}}}D} + \EXP{\be, \lrb{\rmd + \L_{\rmd \bs{\chi}}}T} + \EXP{\gamma, \lrb{\rmd + \L_{\rmd \bs{\chi}}}S} \right. \\
    &\quad\left.- \underbrace{\sum_{i=1}^{N_\xi} h_i\lrb{\B{m}^{(h)}, \bs{\xi}_i^{(h)}}\circ \rmd W_t^i}_{\text{SALT}} \right.  - \underbrace{\int \EXP{\delta \rmd\bs{\chi}, \B{F}}}_{\text{SFLT}} \, .%+ \EXP{\delta \al, \rmd F_D} +  \EXP{\delta \be, \rmd F_T} + \EXP{\delta \gamma, \rmd F_S} }_{\text{SFLT}}\, ;
\end{split}
}
% \algn{
% \begin{split} \label{eq:action_SALT_SFLT}
%     0 = \delta \mathcal{S} = \delta &\left. \int \rmd \bs{\chi}\cdot\B{m}\,d^3x - h(\B{m}^h,D,T,S)\rmd t - w m_3\,d^3x\,\rmd t + \rmd P(D-1)\,d^3x \right.\\
%     &\quad\left. + \int \al \lrb{\rmd D + \nabla\cdot\lrb{D\,\rmd \bs{\chi}}}\,d^3x + \be \lrb{\rmd T + \rmd \bs{\chi}\cdot \nabla T}\,d^3x + \gamma \lrb{\rmd S + \rmd \bs{\chi}\cdot \nabla S}\,d^3x \right. \\
%     &\quad\left.- \underbrace{\sum_i h_i(\B{m})\circ \rmd W_t^i}_{\text{SALT}} \right.  - \underbrace{\int \delta \rmd\bs{\chi}\cdot \B{F} d^3x}_{\text{SFLT}} \, .%+ \EXP{\delta \al, \rmd F_D} +  \EXP{\delta \be, \rmd F_T} + \EXP{\delta \gamma, \rmd F_S} }_{\text{SFLT}}\, ;
% \end{split}
% } 
The first two lines of \cref{eq:action_SALT_SFLT} are what would be included in the unmodified variational principle. $\rmd \zeta$ is a Lagrange multiplier, enforcing $m_3=0$ and after taking variations can be interpreted as the vertical component of the stochastic transport velocity. Indeed, we may expand $\rmd \zeta = w\rmd t + \sum_i \xi_i^{(z)} \circ \rmd W_t^i$; note that here $\rmd \zeta$ is varied and so the third component of $\bs{\xi}_i$ is treated as a variable in the action, whereas the horizontal components are treated as fixed parameters. The final term on the top line enforces incompressibility, and the Lagrange multiplier $\rmd P$ must be stochastic since a semi-martingale Lagrange multiplier is required to enforce a condition on the semi-martingale $D$, \citep[see ][]{street2021}. On the second line the quantities $\al,\be,\gamma$ are Lagrange multipliers enforcing the fact that $D,T,S$ are advected quantities. The final line contains the modifications required to include SALT or SFLT; we shall not in practice use both SALT and SFLT together, but for compactness of the presentation we include them together here. The first modification, giving SALT, consists of a sum of $N_\xi$ Hamiltonians multiplied by Stratonovich noise. The second, additional term is a Lagrange-d'Alembert term which introduces a shift $\B{F}$ in the momentum. We remark that by including further Lagrange-d'Alembert terms such as $\EXP{\delta\al, \rmd F_D}$ or $\EXP{\delta\be, \rmd F_T}$ etc. we may add arbitrary forcings to the right-hand side of the equations for the advected tracers. However, we do not consider this here. \\
\par 
The equations resulting from the variational principle $\delta \mathcal{S} = 0$ are:
\eqsys{
\delta\B{m}^h &:\qquad     \rmd \bs{\chi}^{(h)} = \dD{h}{\B{m}^h}\rmd t + \sum_{i}\dD{h_i}{\B{m}^h}\circ \rmd W_{t}^{i} \, ;\\
\delta m_3&:\qquad \rmd \chi^{(z)} =  \rmd \zeta \,;\\ %= \dD{\rmd H}{\B{m}} \\
\delta\rmd\bs{\chi} &:\qquad      \B{m}  = \al\diamond D + \beta\diamond T + \gamma\diamond S  + \B{F}  \,; \label{eq:SALT_momentum}\\
 \delta\al &:\qquad     \lrb{\rmd + \L_{\rmd \bs{\chi}}}D  = 0 \,;\label{eq:SALT_density}\\ %\rmd F_D  \\
  \delta\be &:\qquad    \lrb{\rmd + \L_{\rmd \bs{\chi}}}T  = 0 \,;\label{eq:SALT_Temp} \\ %\rmd F_T\\
  \delta\gamma &:\qquad     \lrb{\rmd + \L_{\rmd \bs{\chi}}}S  = 0 \,;\label{eq:SALT_Sal}\\ %\rmd F_S \\
  \delta D &:\qquad    \lrb{\rmd + \L_{\rmd \bs{\chi}}}\al  = -\lrb{\rmd P + \dD{h}{D}\rmd t  } \,; \\
  \delta T &:\qquad    \lrb{\rmd + \L_{\rmd \bs{\chi}}}\be  = -\dD{h}{T}\rmd t\,;  \\
  \delta S &:\qquad    \lrb{\rmd + \L_{\rmd \bs{\chi}}}\gamma  = -\dD{h}{S}\rmd t\,;\\
  \delta \rmd P &:\qquad    D  = 1\,; \\
  \delta \rmd \zeta &:\qquad   m_3  = 0\,;
}

The diamond in \cref{eq:SALT_momentum} is a binary operator acting on two variables that are dual with respect to the inner product $\EXP{\cdot,\cdot}$ (e.g. or scalar and density) and giving a 1-form density. Explicitly, for two dual variables $p,q$ and an arbitrary vector field $X:$ the diamond is defined by the relation $\EXP{p\diamond q, X} = -\EXP{p, \L_{X} q}$. We can compute these explicitly as follows:
\algn{
\dD{h}{D}\diamond D = D\nabla\dD{h}{D} \, ,\qquad \dD{h}{T}\diamond T = -\dD{h}{T}\nabla T \, , \qquad \dD{h}{S}\diamond S = -\dD{h}{S}\nabla S \, .
}
We note that the form of $\rmd \bs{\chi}$ as given in \cref{eq:chi_def} is not an input to the variational principle, but a consequence of it. %{\color{blue} RH: should it be $\rmd \bs{\chi}$? SP: Yes. Corrected.}. 
Indeed, we obtain \cref{eq:chi_def} by defining $\B{v}:= \lrb{\dD{h}{\B{m}^h}, w}$ and $\bs{\xi}_{i} := \lrb{\dD{h}{\B{m}^h}, \xi_i^{(z)}}$. The horizontal velocity is therefore $\B{u} = \dD{h}{\B{m}^h} = \frac{\B{m}^h}{D} - \B{R}$. The fact that $D=1$, combined with \cref{eq:SALT_density} gives the incompressibility condition $\nabla \cdot \rmd \bs{\chi} = \nabla^{(h)}\cdot \rmd\bs{\chi}^{(h)} + \PD{}{z}\rmd\zeta = 0$. By Doob-Meyer decomposition \citep{doob1953stochastic,meyer1962decomposition,meyer1963decomposition}, we can split the incompressibility condition into its drift part and stochastic oscillations. Thus we are able to compute $w,\xi_i^{(z)}$ in terms of $\B{u}$ and $\bs{\xi}^{(h)}$ respectively:
\algn{
    \nabla^{(h)} \cdot \B{u} + \PD{w}{z}  =0 \, ,\qquad  \nabla^{(h)} \cdot \bs{\xi}_i^{(h)} + \PD{\xi_i^{(z)}}{z}  = 0  \, . \label{eq:SALT_SFLT_incompressibility}
}
Boundary conditions at $z=0$ allow us to integrate \cref{eq:SALT_SFLT_incompressibility} in the vertical direction.
To obtain the momentum equation we apply $\lrb{\rmd + \L_{\rmd \bs{\chi}}}$ to both sides of \cref{eq:SALT_momentum} and use the fact that the Lie derivative obeys a Leibniz rule with respect to the diamond operator. After some re-arranging, we obtain:
\algn{
\begin{split} \label{eq:SALT_momentum_eq}
  %      \lrb{\rmd + \L_{\rmd\bs{\chi}}}\lrb{\B{m}^h-\B{F}} & = -\lrb{\rmd P + \dD{h}{D}\rmd t}\diamond D -\dD{h}{T} \diamond T \rmd t  -\dD{h}{S}\diamond S \rmd t  \\
        \lrb{\rmd + \L_{\rmd\bs{\chi}}}\lrb{\frac{\B{m}^h-\B{F}}{D} \cdot d\B{x}}& = - d\lrb{\lrb{\dD{h}{D} - V}\rmd t +\rmd P} + \PD{V}{z}dz \rmd t \,.
\end{split}
}
We shall show in \cref{sec:energy} that the SFLT terms will conserve energy if we require that the momentum shift $\B{F}$ takes a particular form, which is that it satisfies $\lrb{\rmd +\L_{\rmd \bs{\chi}}} \B{F} = \L_{\B{v}} \rmd \Phi$, for some stochastic process $\rmd \Phi$. In this work, we shall assume further that $\rmd \Phi$ has the form $\rmd \Phi = \sum_I \bs{\phi}_I \circ \rmd B_t^I$ for some spatially dependent parameters $\bs{\phi}_I$ and with $B_t^I$ being a set of independent Brownian motions. Because the momentum $\B{m} = (\B{m}^h,0)$ has only horizontal components, we shall assume that $\bs{\phi}_I$ also have only horizontal components. Moreover, we can expand the pressure in terms of its drift component and Brownian increments: $\rmd P = p\rmd t + \sum_i p_i  \circ \rmd W_t^i + \sum_I p_I  \circ \rmd B_t^I$. Thus, writing $\B{m} = \B{u} + \B{R}$ and expanding $\rmd\bs{\chi}$ in terms of $\B{v}$ and $\bs{\xi}_i$, we find that \cref{eq:SALT_momentum_eq} becomes:
\algn{
    \begin{split}\label{eq:SALT_SFLT_momentum}
        \rmd \B{u} &+ \lrsq{\nabla\cdot\lrb{\B{v}\B{u}} + f\hat{\B{z}}\times \B{v} + \nabla p + g(1+b)\hat{\B{z}}}\rmd t \\
        & + \sum_i\lrsq{\nabla\cdot\lrb{\bs{\xi}_i \B{u}} + f\hat{\B{z}}\times \bs{\xi}_i + \nabla\bs{\xi}_i \cdot \B{u} + \nabla \lrb{p_i + \bs{\xi}_i\cdot\B{R}}}\circ  \rmd W_t^i \\
        & -  \sum_I\lrsq{\nabla\cdot\lrb{\B{v}\bs{\phi}_I}   - \nabla\bs{\phi}_I\cdot\B{v}- \nabla\lrb{ p_I - \B{v}\cdot \bs{\phi}_I}}\circ  \rmd B_t^I =0  \, .
    \end{split} 
}
The first line of \cref{eq:SALT_SFLT_momentum} contains the terms of the deterministic momentum equation, the second line contains the SALT terms and the final line contains the SFLT contributions. \Cref{eq:SALT_SFLT_momentum} is a three-dimensional equation, but the third component is the (diagnostic) hydrostatic balance condition rather than a prognostic evolution equation for $w$. In the cases of SALT and SFLT hydrostatic balance includes additional constraints on the stochastic parts of the pressure $\rmd P$:
\algn{\label{eq:SALT_SFLT_pressure}
    \PD{p}{z} & = - g(1+b) \, ,\qquad \PD{p_i'}{z}  = -\PD{\bs{\xi}_i}{z}\cdot\B{u}\, , \qquad    \PD{p_I'}{z}  = - \PD{\bs{\phi}_I}{z}\cdot\B{v}  \, ,
}
where we have the definitions of the shifted stochastic pressure terms $p_i' := p_i + \bs{\xi}_i\cdot\B{R}$ and $p_I' := p_I - \B{v}\cdot\bs{\phi}_I$. We solve \cref{eq:SALT_SFLT_pressure} by imposing the following surface pressure boundary conditions:
\algn{\label{eq:SALT_SFLT_pressure_BCs}
    p|_{z=0} & = g \eta\,,  \qquad
    p_i'|_{z=0}  = \psi_i\,, \qquad
    p_I'|_{z=0}  = \psi_I\,,
}
where $\eta$ is the free surface height. The boundary condition on $p$ is that used in the linear free surface approximation, which is employed in FESOM2 \citep{danilov_sidorenko_wang_jung_2016}. $\psi_i$ and $\psi_I$ are functions only of the horizontal direction and are arbitrary. They may be used to introduce some stochastic atmospheric forcing at the ocean surface, but  we do not consider this in the present work. For simplicity we shall set $\psi_i = \psi_I = 0$ for all $i,I$. Solving \cref{eq:SALT_SFLT_pressure} with the boundary conditions in \cref{eq:SALT_SFLT_pressure_BCs} gives us the following:
\begin{subequations}
\algn{
    p &= g(\eta-z) + g\int_z^0 b dz'\,,\\
    \quad  p_i' &=  \psi_i + \int_z^0 \PD{\bs{\xi}_i}{z}\cdot\B{u} dz'\,, \quad \\
    p_I' &= \psi_I + \int_z^0\PD{\bs{\phi}_I}{z}\cdot\B{v} dz'\,.
}
\end{subequations}
A more exact condition on the deterministic pressure would be $p|_{z=\eta}=0$. Using this gives almost the same result for $p$ except that the upper limit of the integral will instead be $\eta$. \\
\par
The equation for the evolution of the free surface height $\eta$ is obtained by integrating the incompressibility condition and using appropriate surface boundary conditions. For the linear free surface approximation we take $w|_{z=0}\rmd t = \rmd \eta$; at the bottom boundary $z=-H(x,y)$ we have $\rmd \bs{\chi}|_{z=-H}\cdot\nabla\lrb{z+H} = 0$. Thus, integrating the incompressibility condition in the vertical direction from $z=-H$ to $z=0$ we find, in the linear free surface case:
\algn{\label{eq:SALT_SFLT_linfs}
\rmd \eta   +   \nabla\cdot\int_{-H}^{0}  \B{u} \rmd t\,dz & = 0\,.
}
Again, the more exact boundary condition would be $\rmd \bs{\chi}|_{z=\eta} \cdot \nabla\lrb{z - \eta} = \rmd \eta $ ad in this case \cref{eq:SALT_SFLT_linfs} is modified by $\B{u}\rmd t \arr \B{u}\rmd t + \sum_i \bs{\xi}_i \circ \rmd W_t^i$ and the upper limit of the integral will be $\eta$ rather than $0$. However, for our numerical simulations we use the linear free surface.\\
\par
From \cref{eq:SALT_Temp,eq:SALT_Sal} we have the advection equations:
\algn{      
\rmd T &+ \B{v}\cdot\nabla T\rmd t + \sum_i \bs{\xi}_i\cdot\nabla T \circ \rmd W_t^i = 0  \label{eq:SALT_SFLT_T}\,,\\
 \rmd S &+ \B{v}\cdot\nabla S\rmd t + \sum_i \bs{\xi}_i\cdot\nabla S \circ \rmd W_t^i = 0\label{eq:SALT_SFLT_S}\,,
}
for temperature and salinity respectively. The horizontal component of the momentum equation \cref{eq:SALT_SFLT_momentum}, along with the solutions \cref{eq:SALT_SFLT_pressure} for pressure (with the equation of state $b=b(T,S,z)$), the incompressibility conditions \cref{eq:SALT_SFLT_incompressibility}, the tracer advection equations \cref{eq:SALT_SFLT_T,eq:SALT_SFLT_S} and the linear free surface equation \cref{eq:SALT_SFLT_linfs} give us a complete set of fluid equations, the Primitive Equations with SALT and SFLT.

\subsection{Conservation Laws}
\label{sec:energy}
The key benefit of the SALT and SFLT frameworks is that they retain some of the fundamental conservation properties possessed by the deterministic equations. By writing the Primitive Equations in the geometric form given in \cref{eq:SALT_momentum_eq,eq:SALT_density,eq:SALT_Temp,eq:SALT_Sal}, we may demonstrate the effect of the stochastic frameworks on these conservation laws. First, we consider energy conservation. The total energy is equal to the Hamiltonian, as given in \cref{eq:hamiltonian}. For convenience of notation, we define $\tl{h}(\B{m}, D, T, S, w) = h(\B{m}^h, D, T, S) + \EXP{m_3,w}$. $h$ and $\tl{h}$ are equal on solutions of the equations, but we have $\dD{\tl{h}}{\B{m}} = \B{v}$. By direct calculation, the time evolution of the energy is given by:
\algn{
\begin{split}
    \rmd h  &= \sum_i \lrsq{\EXP{\dD{\tl{h}}{\B{m}}\,,\, \sum_i \L_{\bs{\xi}_i}\B{m}^h } + \EXP{g(1+b), \xi_i^{(z)}}} \circ \rmd W^i_t  \\
    & \qquad\qquad  - \EXP{\dD{\tl{h}}{\B{m}}, (\rmd + \L_{\rmd\bs{\chi}})\B{F} }\,. \label{eq:energy_cons_stoch}
\end{split}
}
Thus, the energy conservation property is violated by the stochastic terms. The two terms on the right-hand side of the pairing in \cref{eq:energy_cons_stoch} come from SALT and SFLT respectively. However, as shown in \citep{HolmHu2021}, the energy deviation from SFLT can be nullified by choosing $\lrb{\rmd +\L_{\rmd \bs{\chi}}} \B{F} = \sum_I \L_{\B{v}} \bs{\phi}_I \circ \rmd B_t^I$ for some parameters $\bs{\phi}_I(\B{x})$. Indeed, by the anti-symmetry of the vector field commutator:
\algn{
    \EXP{\dD{\tl{h}}{\B{m}}\,,\, \sum_I\L_{\B{v}} \bs{\phi}_I \circ \rmd B_t^I} = \EXP{\lrsq{\dD{\tl{h}}{\B{m}}\,,\, \B{v}}\,,\, \sum_I \bs{\phi}_I \circ \rmd B_t^I} = 0\,,
}
where the square bracket $[\cdot]$, denotes the commutator of vector fields. Thus, energy conservation is broken by SALT but preserved by a class of stochastic forcing in SFLT. In the remainder of the paper, we shall assume the stochasticity introduced by SFLT are in the energy preserving form. \\
\par
The next conservation law we consider is the Kelvin circulation theorem. The evolution of the circulation corresponding to  \cref{eq:SALT_momentum_eq} is given by:
\algn{
    \rmd \oint_{C(t)} \frac{\B{m}^h}{D}  \cdot d\B{x} & = -g\oint_{C(t)} b(T,S,z)dz \,\rmd t + \sum_I\oint_{C(t)} (\curl\bs{\phi}_I \times\B{v})\cdot d\B{x}\,  \circ \rmd B^I_t\,,% \sum_I\oint_{C(t)}\frac{1}{D}\lrb{ \curl\bs{\phi}_I \times\B{v}+\nabla\lrb{\B{v}\cdot \bs{\phi}_I}}\cdot d\B{x}\circ \rmd B^I_t\,.
}
where $C(t)$ is a closed loop moving with the transport velocity $\rmd \bs{\chi}$. We see that SALT affects the circulation theorem only by modifying the advection of the loop; thus the circulation theorem for SALT  is the same as in the deterministic case, but with the circulation considered around a stochastically-transported loop. Therefore, circulation is generated only by buoyancy gradients being misaligned with the vertical direction. In SFLT, on the other hand, there are additional forces introduced, which generate the circulation of fluid momentum.\\
\par
The evolution of potential vorticity associated with \cref{eq:SALT_momentum_eq} can be expressed as
\algn{
     \lrb{\rmd + \rmd \bs{\chi}\cdot\nabla } q &= \frac{1}{D}\bs{\omega}\cdot \nabla\lrb{ \PD{b}{z}\rmd \chi^{(z)} } + \frac{1}{D}\nabla b\cdot \sum_I \lrsq{ \nabla\cdot\lrb{\B{v} \bs{\omega}_I} - \bs{\omega}_I\cdot\nabla\B{v} }\circ \rmd B^I_t\,,
}
where $\bs{\omega}:=\curl\lrb{\B{m}^h/D}$ is the relative vorticity, $\bs{\omega}_I = \curl \bs{\phi}_I$ is the stochastic vorticity generated by SFLT and  $q:= \frac{1}{D} \bs{\omega} \cdot \nabla b$ is the potential vorticity. Similar to the Kelvin circulation theorem, SALT introduces stochasticity in the transport velocity $\rmd \bs{\chi}$, while SFLT introduces stochastic forces that act on the advection of fluid potential vorticity. If we assume that the buoyancy has no explicit dependence on the vertical coordinate, i.e. $\PD{b}{z}=0$, then $q$ is purely advected by the flow in the absence of SFLT.

\section{Calibration of the stochastic parameters}
\label{sec:Xis}
\subsection{Lagrangian Paths}\label{sec:Lagrangian paths}
In order to calibrate the parameters $\bs{\xi}_i$ used in SALT we propose to use the method of Lagrangian paths introduced in \citep{cotter2019numerically,cotter2018modelling}. \par First, we perform a fine-grid model run, which we shall take to be the `truth'. Resulting from this run we get an output velocity $\B{v}(\B{x},t)$ saved at times $t\in\lrc{t_1,...,t_{N-1}, t_N}$, where the time interval between subsequent sample times, $t_{i+1}-t_i$, is greater than the velocity decorrelation time, defined to be the smallest $\tau$ at which the auto-correlation function $C(\tau)$ % = \frac{\EXP{\B{u}'(t+\tau)\cdot\B{u}'(t)}}{\EXP{\B{u}'(t)\cdot\B{u}'(t)}}
is less than $e^{-1}$.
Suppose the fine-grid resolution is $M$ times that of the coarse grid, in which case the coarse-grid time step is given by $\Delta t_c := M\Delta t_f$, where $\Delta t_f$ is the time step for the fine-grid model run. In order to compute Lagrangian paths we also save $\B{v}(\B{x},t)$ at $t\in\lrc{t_i, t_{i}+\Delta t_f, ..., t_i + (M-1)\Delta t_f}$ for each $i=1,...,{N}$. \par
% To obtain a coarse-grid velocity we apply to $\B{v}_t$ a smoothing operator $\B{v}_t \mapsto \bar{\B{v}}_t$, such that $\bar{\B{v}}_t$ is defined on the coarse grid. 
To obtain the corresponding coarse-grid velocity $\bar{\B{v}}(\bar{\B{x}},t)$ from $\B{v}(\B{x},t)$, we apply a coarse-graining operator to $\B{v}(\B{x},t)$, which consists of a local average over fine-grid points, to obtain a velocity $\bar{\B{v}}(\bar{\B{x}},t)$ defined on the coarse grid. Considering a distribution of tracer particles whose initial positions $\B{x}_0^r$ are the (three-dimensional) coordinates of the coarse-grid nodes (enumerated by $r$), we compute Lagrangian paths on the fine grid and coarse grid respectively:
%Now we can compute Lagrangian paths using the velocities $\B{v}_t$ and $\bar{\B{v}}_t$. We take as initial particle positions the coordinates of the coarse-grid nodes, and we compute two Lagrangian paths: one from the fine-grid velocity, integrated for $M$ fine-grid time steps; and one from the coarsened velocity, integrated for one coarse-grid time step:
\eqsys{
    \B{x}_f^r \lrb{t_i+M\Delta t_f} &:= \B{x}_0^r + \sum_{m=0}^{M-1} \B{v}\lrb{\B{x}_f^r \lrb{t_i+m\Delta t_f}, t_i+m\Delta t_f} \Delta t_f\,,\\
    \B{x}_c^r \lrb{t_i+M\Delta t_f} &:= \B{x}_0^r + \bar{\B{v}}\lrb{\B{x}_c^r(t_i), t_i} \Delta t_c\,,
}
where $\B{x}_r^f\lrb{t_i+M\Delta t_f} $ and $\B{x}_c^r\lrb{t_i+M\Delta t_f}$ are the Lagrangian paths computed as integral curves of $\B{v}_f$ and $\bar{\B{v}}$ respectively; the integral is carried out over one coarse-grid time-step, which is equivalent to  $M$ fine-grid time steps. We can then define the difference $\Delta \B{x}_{r,i} = \Delta \B{x}(t_i, \B{x}_0^r):= \B{x}_f^r \lrb{t_r+M\Delta t_f} - \B{x}^c \lrb{t_r+M\Delta t_f}$ and apply the method of \citep{Hannachi_2007} to compute the Empirical Orthogonal Functions (EOFs). To summarise, we subtract off the time mean to define $\Delta \B{x}_{r,i}' :=\Delta \B{x}_{r,i} - \frac{1}{N}\sum_{i=0}^{N-1}\Delta \B{x}_{r,i}$. In the $x$-direction we then have a matrix with components $\Delta x_{r,i}'$. From this we construct the matrix $\Lambda^{(x)}$ which has components $\Lambda_{rs}^{(x)} = \frac{1}{N}\sum_{i=0}^{N-1}\Delta x_{r,i}' \Delta x_{s,i}' $. The EOFs in the $x$-direction are then defined to be the eigenvectors of the matrix $\Lambda^{(x)}$ which we denote as $a^{(x)}_{i}$, for $i=1\ldots N$. They are normalised in the sense that $\sum_{\B{x}}a_i^{(x)}(\B{x})a_j^{(x)}(\B{x}) = \delta_{ij}$, where the sum is over all grid points. We apply the same process to the $y$-component $\Delta y_{r,i}'$ to obtain $N$ eigenvectors in the $y$-direction, which we denote $a^{(y)}_{i}$. We do not compute the eigenvectors for the $z$-direction since these will be obtained from the incompressibility condition.\\
\par
We remark that the method we have used here, in which we compute the EOFs of each component of $\Delta \B{x}$ separately, is different from the method found in other sources \citep[e.g.][]{holmes_lumley_berkooz_1996}, in which the components are computed together and we obtain a set of two-component eigenvectors $\B{a}_i$ immediately, with one eigenvalue $\lambda_i$ corresponding to each of these EOFs. However, this method was attempted for SALT runs in the current set-up and the results of model runs were less successful. For this reason we have chosen to compute the components separately.\\
%{\color{blue}RH: shall we throw in a remark about using EOF on the vector of horizontal velocities?}
\par Thus, in our case we have $N$ eigenvectors in each of the horizontal directions and these will have associated eigenvalues $\lambda_i^{(x)}$ and $\lambda_i^{(y)}$. We define the horizontal components of $\bs{\xi}_i$ by a re-scaling of these eigenvectors. The magnitude of the eigenvalue $\lambda_i^{(x)}$ gives an indication of how much of the variance is captured by the corresponding eigenvector. Therefore, we choose to scale the parameters so that $\EXP{\bs{\xi}_i^{(h)}, \bs{\xi}_i^{(h)}} \propto \lambda_i$. Moreover, in order to ensure that the different methods for computing $\bs{\xi}_i$ may be compared fairly, we require that the $L_2$-norm of the sum be the same for each method. Thus we impose the following:
\algn{
    \frac{1}{V_{tot}}\sum_{i=1}^{N_{\xi}}\EXP{\bs{\xi}_i^{(h)}, \bs{\xi}_i^{(h)} } = \gamma^2 
}
where $\gamma$ is a constant with units $ms^{-1/2}$, which we shall choose later; $V_{tot}$ is the total volume of the domain. and $N_{\xi}\leq N$ is the number of EOFs we choose to keep for our model runs. The total integral, denoted by angle brackets, is defined by $\EXP{\B{a},\B{b}}:=\sum_{\B{x}} \B{a}(\B{x})\cdot\B{b}(\B{x})V(\B{x})$. We can achieve the required properties by choosing the following scaling:
\algn{
    \xi_i^{(x)}(\B{x}) = \gamma \sqrt{\frac{\lam_i^{(x)}}{\lam_{tot}}\cdot \frac{V_{tot}}{V(\B{x})}} a_i^{(x)}(\B{x})
}
where $V(\B{x})$ is the volume of the grid cell located at $\B{x}$ and we have defined $\lam_{tot} := \sum_{i=1}^{N_{\xi}}(\lam_i^{(x)} + \lam_i^{(y)})$. 
 After computing the horizontal components in this way, $\bs{\xi}_i^{(h)}$ are then smoothed to zero near the boundaries in order to enforce the impermeability condition at the boundary, $\bs{\xi}_i^{(h)}\cdot\B{n}=0$, where $\B{n}$ is the normal to the boundary and $\bs{\xi}_i^{(h)} = (\xi^{(x)}_i,\xi^{(y)}_i)$ is the horizontal part of $\bs{\xi}_i = (\xi^{(x)}_i,\xi^{(y)}_i, \xi_i^{(z)})$. \\
 \par
For the $z$-component we use the incompressibility condition \cref{eq:SALT_SFLT_incompressibility} along with the impermeability condition $\bs{\xi}_{i}\cdot\nabla \lrb{z+H}=0$ at the lower boundary $z=-H$ to obtain:
\algn{
   \xi_i^{(z)}  & = - \nabla^{(h)}\cdot  \int_{-H}^z  \bs{\xi}_i^{(h)} dz\,,\label{eq:calibrate xi z}
}
where $\nabla^{(h)} = (\PD{}{x}, \PD{}{y})$ is the horizontal gradient. This method for computing the vertical component of $\bs{\xi}_i$ is applicable to any system of fluid equations with an incompressibility condition.
We could, alternatively, compute all three components of $\bs{\xi}_i$ as EOFs of the three components of $\Delta \B{x}$. However, the resulting three-component vector $\bs{\xi}_i$ will not be guaranteed to be divergence-free. We would then need to subtract off the divergent part $\bs{\xi}_i \arr \bs{\xi}_i' =  \bs{\xi}_i - \nabla \Delta^{-1}\lrb{\nabla\cdot\bs{\xi}_i}$ where $\Delta^{-1}$ is the inverse Laplacian.
However, computing the divergent part of the vector $\bs{\xi}_i$ is computationally expensive; moreover, the components of $\bs{\xi}_i$ computed in this way will not be guaranteed to be orthogonal with respect to $\EXP{\cdot,\cdot}$.  Thus in this paper we consider only the $\bs{\xi}_i$ for which the vertical components are computed from integrating the incompressibility condition.
% However, this is more computationally expensive and so here we stick to computing the vertical component from integrating the incompressibility condition.

\subsection{Eulerian Differences}
To calibrate the parameters $\bs{\phi}_I$ used in SFLT we propose an alternative method by using differences in fixed Eulerian coordinates. Consider the deterministic momentum equation given by:
\algn{
\begin{split}
        \lrb{\rmd + \L_{\B{v}\rmd t}}\lrb{\B{m}^h} & = -\lrb{p + \dD{h}{D}}\diamond D\rmd t -\dD{h}{T} \diamond T \rmd t  -\dD{h}{S}\diamond S \rmd t\,,
\end{split}
}
and the SFLT equation:
\algn{
\begin{split}
        \lrb{\rmd + \L_{\bar{\B{v}} \rmd t}}\lrb{\bar{\B{m}}^h} - \sum_I\L_{\bar{\B{v}}}\lrb{\rmd \Phi} = -\lrb{p + \dD{h}{\bar{D}}}\diamond \bar{D}\rmd t -\dD{h}{\bar{T}} \diamond \bar{T} \rmd t  -\dD{h}{\bar{S}}\diamond \bar{S} \rmd t \,,
\end{split}
}
where the notation $\bar{(\cdot)}$ are used on the variables of the SFLT equations to emphasise the difference between deterministic and stochastic variables. The goal of the stochastic parameterisation is to decompose the ``true'' fluid flow to a slow drift component and a rapid fluctuating component whose amplitude can be estimated from data. In the example of estimating the momentum fluctuation $\rmd \Phi$ of $\bar{\B{m}}^h$, we denote the slow drift component as $\bar{\B{m}}^h$ and we seek the solution to the minimisation problem
\algn{
    \min_{\rmd \Phi} \mathbb{E}\lrsq{\rmd \B{m}^h - \rmd \bar{\B{m}}^h}\,.
}
Assuming $D, T$ and $S$ do not have rapidly fluctuating components, the minimisation problem becomes
\algn{
    \min_{\rmd \Phi} \mathbb{E}\lrsq{\L_{\B{v}}(\B{m}^h\rmd t) - \L_{\bar{\B{v}}}(\bar{\B{m}}^h\rmd t - \rmd \Phi)}\,.
}
We see that this minimisation problem can be solved by taking \\ $\rmd \Phi = \lrb{\B{m}^h- \bar{\B{m}}^h}\rmd t = \lrb{\B{u} - \bar{\B{u}}}\rmd t$. Therefore, we define the differences 
\algn{
\Delta \B{x}_{r,I} := \Delta \B{x}(t_I, \B{x}_0^r) = \lrsq{\B{u}(t_I, \B{x}_0^r) - \bar{\B{u}}(t_I,\B{x}_0^r)}\Delta t_c
}
for $I=1,...,N$. We then assume the expansion $\rmd \Phi = \sum_{I=1}^{N} \bs{\phi}_I \circ \rmd B_t^I$. As before, we subtract the time-mean to obtain $\Delta \B{x}'_{r,I} = \Delta \B{x}_{r,I} - \frac{1}{N}\sum_{I=0}^{N-1}\B{x}_{r,I}$ and then compute the EOFs exactly as we did in \cref{sec:Lagrangian paths} to get our parameters $\bs{\phi}_I$. 
\par
In both methods we initially compute horizontal components of the stochastic parameters using EOFs, but for SALT there is the additional step of integrating the incompressibility condition to obtain the vertical component. The vertical component is not needed for $\bs{\phi}_I$ since it is a part of the decomposition of the fluid momentum $\B{m}$, the vertical part of which vanishes in the Primitive Equations. In fully three-dimensional models in which the vertical component of the momentum is non-zero, the Eulerian differences of the momenta will be a three-dimensional object and one can compute all three components of the parameters $\bs{\phi}_I$ using EOFs. \par
% We do not need a vertical component of $\phi_I$ because we are working with the Primitive equations and therefore the vertical momentum is zero.
% In fully three-dimensional models in which the momentum does have a vertical component it will be necessary to define $\Delta \B{x}$ as a three-dimensional object. \par 
We can also consider using Eulerian differences as an option for $\bs{\xi}_i$ in SALT. This effectively means approximating the fine-grid Lagrangian path by taking only one time-step in the coarse grid: $\B{x}_f \approx \B{v}_f(\B{x}_0, t)\Delta t_c$. We can expect that this will be a reasonable approximation for small $M$, but for larger $M$ the Lagrangian paths method will diverge from the Eulerian differences. In our numerical investigations in SALT we shall consider $\bs{\xi}_i$ computed from both the Lagrangian paths method and Eulerian differences method. For SFLT we also consider $\bs{\phi}_I$ computed from Lagrangian paths (for completeness) as well as those computed by Eulerian differences as described above.

\section{Results}
\label{sec:results}

\begin{figure}
    \centering
    \includegraphics[width=0.9\textwidth]{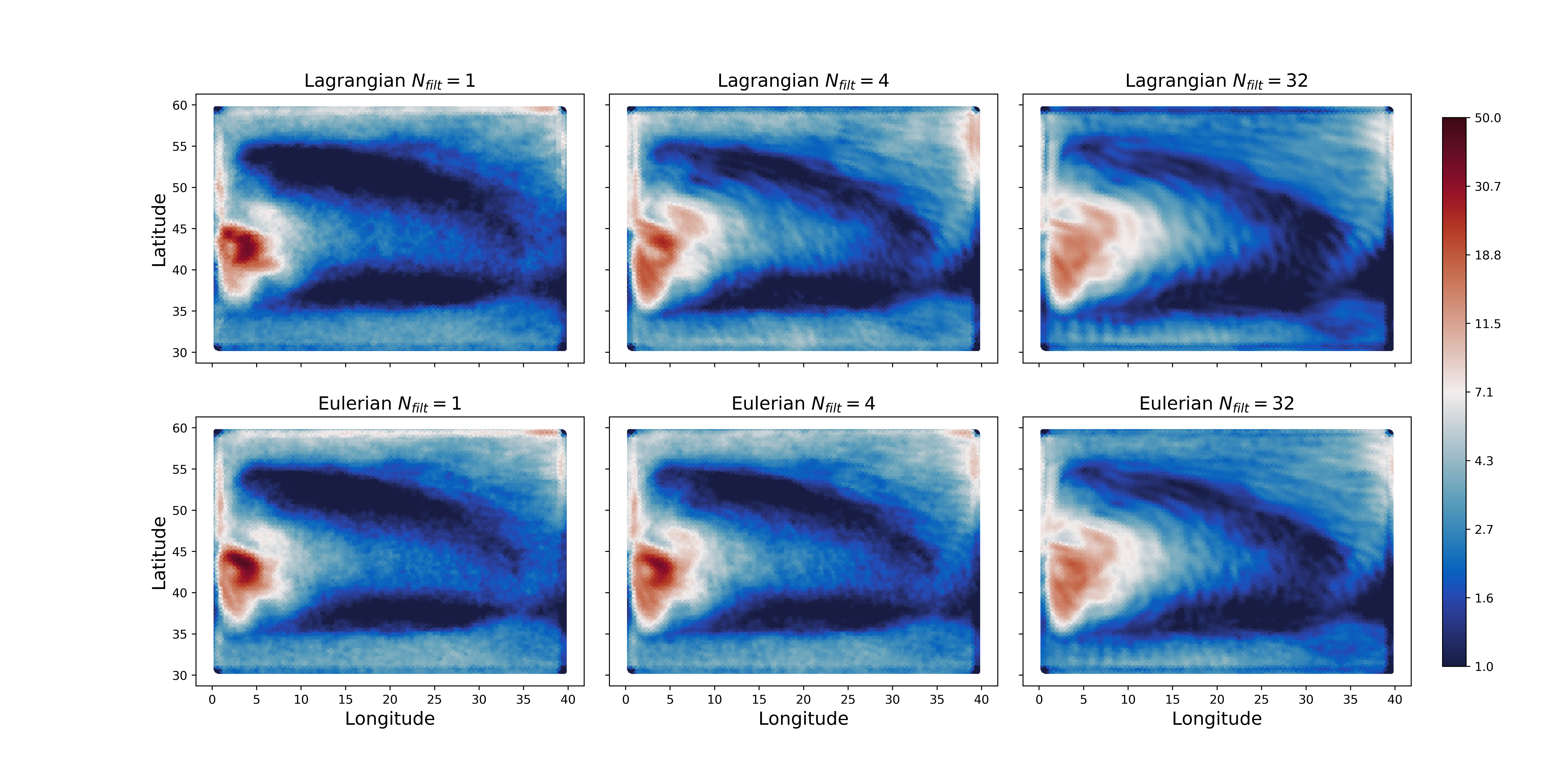}
    \captionsetup{width=0.9\linewidth}
    \caption{$\frac{1}{\gamma}\lrb{\frac{1}{N_\xi}\sum_{i=1}^{N_{\xi}} \bs{\xi}_i^{(h)}\cdot \bs{\xi}_i^{(h)}}^{1/2}$ in the upper fluid layer for different methods of computing $\bs{\xi}_i^{(h)}$. Top row: $\bs{\xi}_i^{(h)}$ computed from Lagrangian paths for different strengths of smoothing filter. Bottom row: $\bs{\xi}_i^{(h)}$ computed from Eulerian differences for different strengths of smoothing filter }
    \label{fig:xis}
\end{figure}

\begin{figure}
    \centering
    \includegraphics[width=0.9\textwidth]{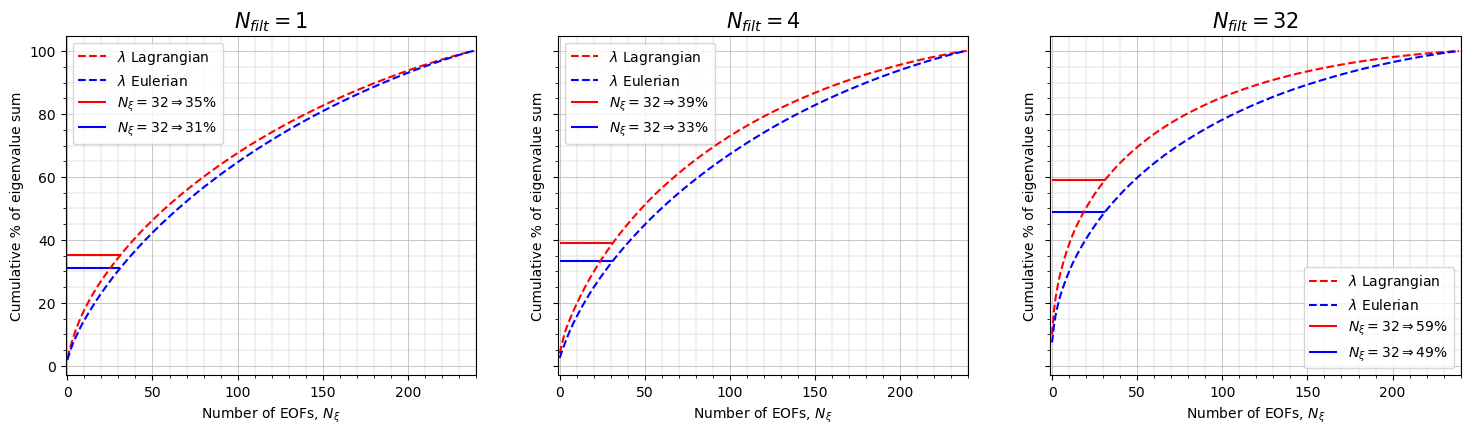}
    \captionsetup{width=0.9\linewidth}
    \caption{Eigenvalue spectra of zonal $\bs{\xi}_i$, plotted for three different values for $N_{filt}$. On each panel is shown the spectrum for the EOFs calculated by Lagrangian trajectories and Eulerian differences. The horizontal lines show what the percentage of the total variance is captured by choosing $N_{\bs{\xi}} = 32$ EOFs.}
    \label{fig:spectra_xi}
\end{figure}

We solve the Primitive Equations using the FESOM2 code on a rectangular domain $[0,40^\circ]\times[30^\circ,60^\circ]\times[0,-H]$, where $H=1600\rm{m}$ is the depth of the domain and the bathymetry is flat. Impermeability conditions are imposed at all boundaries. The model is spun up for three years from zero initial velocity and an initial temperature profile given by $T(z) = T_0 + \frac{\lam}{\al \rho_0} \lrb{  (1-\beta)\tanh\lrb{\frac{z}{z_0}} + \beta \frac{z}{H}}$, which is based on the test case described in \citep{SOMA,sein2016designing}. We take $T_0 = 25^\circ C$, $\be=0.05$, $\lambda = 5 \rm{kgm}^{-3}$, $z_0 = 300\rm{m}$, $\rho_0 = 1030\rm{kg}\rm{m}^{-3}$, and $\al = 0.00025\rm{K}^{-1}$.  For simplicity, salinity is kept constant and we use a linear equation of state which depends only on temperature: $b = - \al(T-10^\circ C)$. The flow is driven by a wind forcing in the upper layer given by $\bs{\tau}(x,y) = \frac{-\tau_0\Delta z_0}{\rho_0}\cos\lrb{\frac{\pi y}{15^\circ}}\hat{\B{x}}$, where $\Delta z_0 = 10\rm{m}$ is the thickness of the upper layer; $\tau_0 = 0.2 \rm{ms}^{-2}$ is the wind strength. The vertical discretisation consists of 23 layers, with layer thicknesses increasing with depth. For the horizontal discretisation we take a fine grid of spacing $1/4^\circ$ and a coarse grid of spacing $1/2^\circ$. At the latitudes we are considering, $1/4^\circ$ corresponds to an eddy-permitting model, while $1/2^\circ$ may be considered non-eddy resolving, \citep[see ][]{hallberg_2013}. %Considering the effect of parameters generated from an eddy-resolving run on an eddy-permitting simulation is beyond the scope of the present work.
We run the deterministic model on the fine grid and the coarse grid, and carry out the SALT and SFLT runs on the coarse grid. All coarse-grid runs are begun from the same initial condition, being the final time snapshot after the three-year spin-up period; the fine-grid run is begun from the end of the three-year spin-up on the fine grid. We save data in each case at intervals of 15 days, over a time period of 10 years, for a total of 240 snapshots. From the fine grid data we have the `truth' velocity $\B{v}_f$. To this we apply a coarse-graining $\bar{\B{v}}$; we then follow the procedures outlined in \cref{sec:Xis} to compute $\bs{\xi}_i^{(h)}$ and $\bs{\phi}_I$. However, there is no canonical choice for how the coarse-graining should be done. We consider a filter defined by an equally-weighted nine-point average over nearest neighbours, and we denote this filter $\mathcal{F}$; this filter, applied once, has a width equal to the spacing on the coarse grid, i.e. $1/2^\circ$. The coarse-graining will then be done by applying this filter $N_{filt}$ times successively, then projecting onto the coarse grid. Thus, the smoothing filter applied $N_{filt}$ times will be denoted $\mathcal{F}^{N_{filt}}$; this has a width ${N_{filt}}/2$ degrees with a stronger weighting for points closer to the centre of the filter. We consider the cases $N_{filt}=1, 4, 32$. \par
From the deterministic model run, we have velocities saved at 240 time snapshots, so we can use these to compute 240 EOFs. We do this for both the Lagrangian paths method and the Eulerian differences method, for each of the three choices of $N_{filt}$; this gives a total of six sets of parameters. In our model runs we shall choose to keep $N_{\xi} = N_{\phi} = 32$ of these parameters for each run. In \cref{fig:xis} we plot the square-root of the sum of the squares of these parameters (before re-scaling by $\gamma$) as a field in space. From \cref{fig:xis} it appears the differences between Lagrangian paths or Eulerian differences are minimal. %For $N_{filt}=1$ there is some difference in the magnitude of the resulting $\bs{\xi}_0$, but the shape of the field is still similar. 
We remark that here the time-steps on the fine and coarse grids differ only by a factor of 2; it is expected that if a bigger difference in resolution is used, then more steps will be needed in computing the Lagrangian paths and therefore the corresponding parameters will differ more substantially. The number of times we apply the smoothing operator, however, has a much greater effect and we see significantly different fields with $N_{filt}=32$ than we do with $N_{filt}=4$ or $N_{filt}=1$. Indeed, it appears from \cref{fig:xis} that the weaker filter causes the parameters to be more strongly concentrated around the western boundary, whereas for the stronger filter the parameters are spread more across the domain.  
\par
The cumulative spectra of the EOFs are shown in \cref{fig:spectra_xi}. These spectra show us how many EOFs are needed to capture a given percentage of the total variability; or conversely, how much variance is captured by a given number of EOFs. We show in each case how much variability is captured by using $32$ EOFs. In all cases the Lagrangian paths method gives a slightly higher variability captured, though the difference is small, especially for the smaller values of $N_{filt}$. A much bigger variability is captured, however, in the $N_{filt}=32$ case when compared with the $N_{filt}=1$ case. 
%\begin{framed}
%{\color{red} RH: what is the number of EOFs used in SALT and SFLT runs? I remember it being 16 but its not mentioned in this section.}
%\end{framed}
\par We implemented SALT and SFLT into FESOM2 (see \cref{sec:Numerics}) and ran the model with each choice of parameters and with the appropriate re-scaling as detailed above. For all SALT runs we use $N_{\xi} = 32$ with the scaling $\gamma = 2\times10^{-3} ms^{-1/2}$.  For SFLT we also take $N_{\phi}=32$ but scale the parameters with $\gamma= 10^2 ms^{-1/2}$. This re-scaling is chosen empirically taking $\gamma$ to be the largest value possible that will not result in model blow-up. It appears that the magnitude of parameters that we are able to use for SFLT is much higher. This is possibly due to the fact that SFLT does not involve any direct modification of the tracer equation. SALT, on the other hand, includes an advection of the temperature by the stochastic transport velocity; using higher values for this velocity may destabilise the tracer equation and cause model blow-up.\\
\par
\begin{figure}
    \centering
    \includegraphics[width=1.0\textwidth]{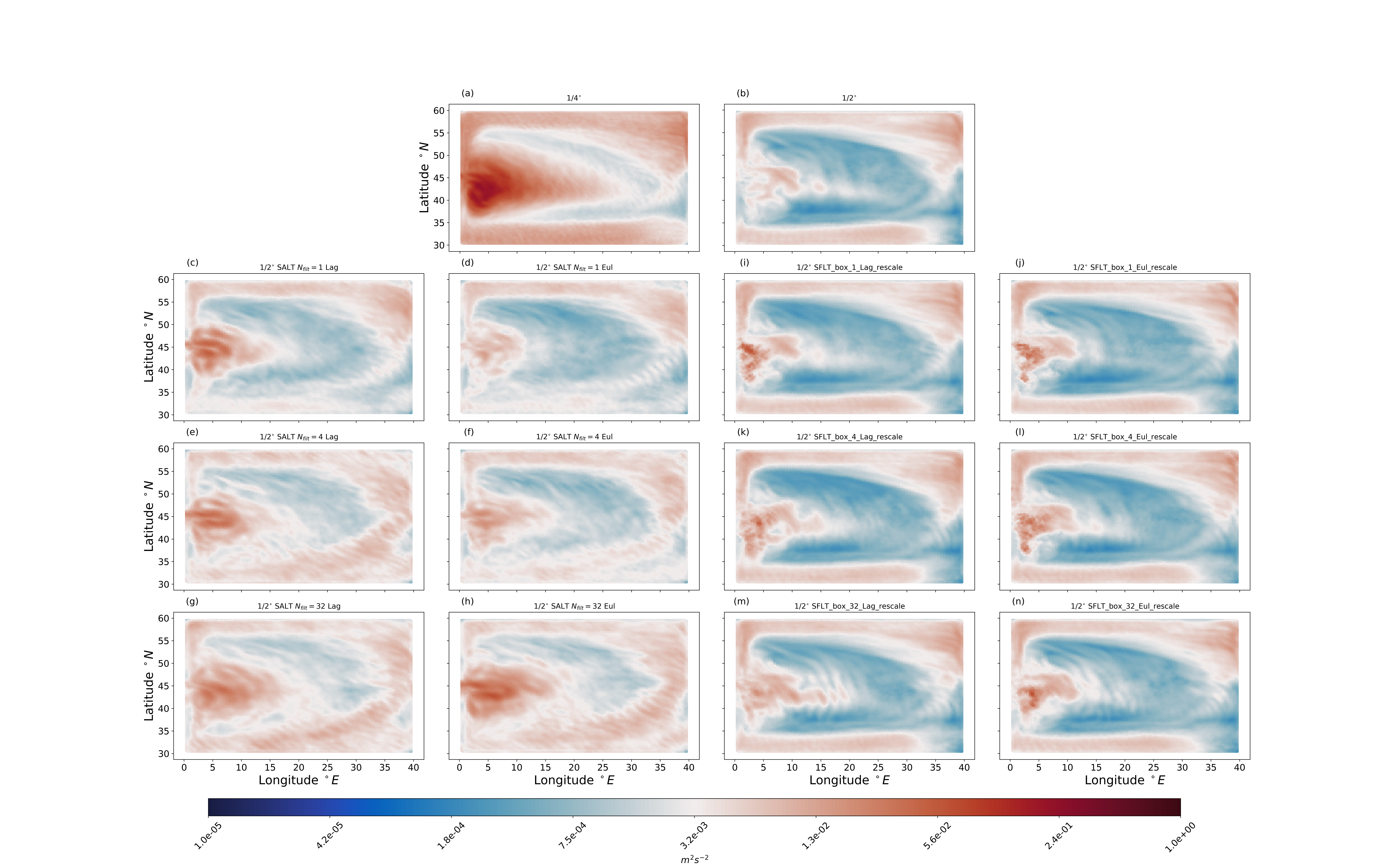}
    \captionsetup{width=0.9\linewidth}
    \caption{Time-average of eddy kinetic energy at depth 16m below the surface. Panel (a) is from the high-resolution ($1/4^\circ$) deterministic model, while (b) is from the low-resolution ($1/2^\circ$) deterministic model. Panels (c), (g), (k) are the results of model runs at $1/2^\circ$ with SALT, where $\bs{\xi}_i$ are computed using Lagrangian differences using a coarse velocity defined by applying the smoothing filter 1, 4 and 32 times respectively. Panels (d), (h), (l) are also SALT runs but $\bs{\xi}_i$ are computed from Eulerian differences rather than Lagrangian trajectories. Panels (e), (i), (m) are SFLT runs with $\bs{\phi}_I$ computed from Lagrangian trajectories, while (f), (j), (n) have $\bs{\phi}_I$ computed from Eulerian differences.  }
    \label{fig:eke_SALT_SFLT}
\end{figure}
\begin{figure}
    \centering
    \includegraphics[width=0.9\textwidth]{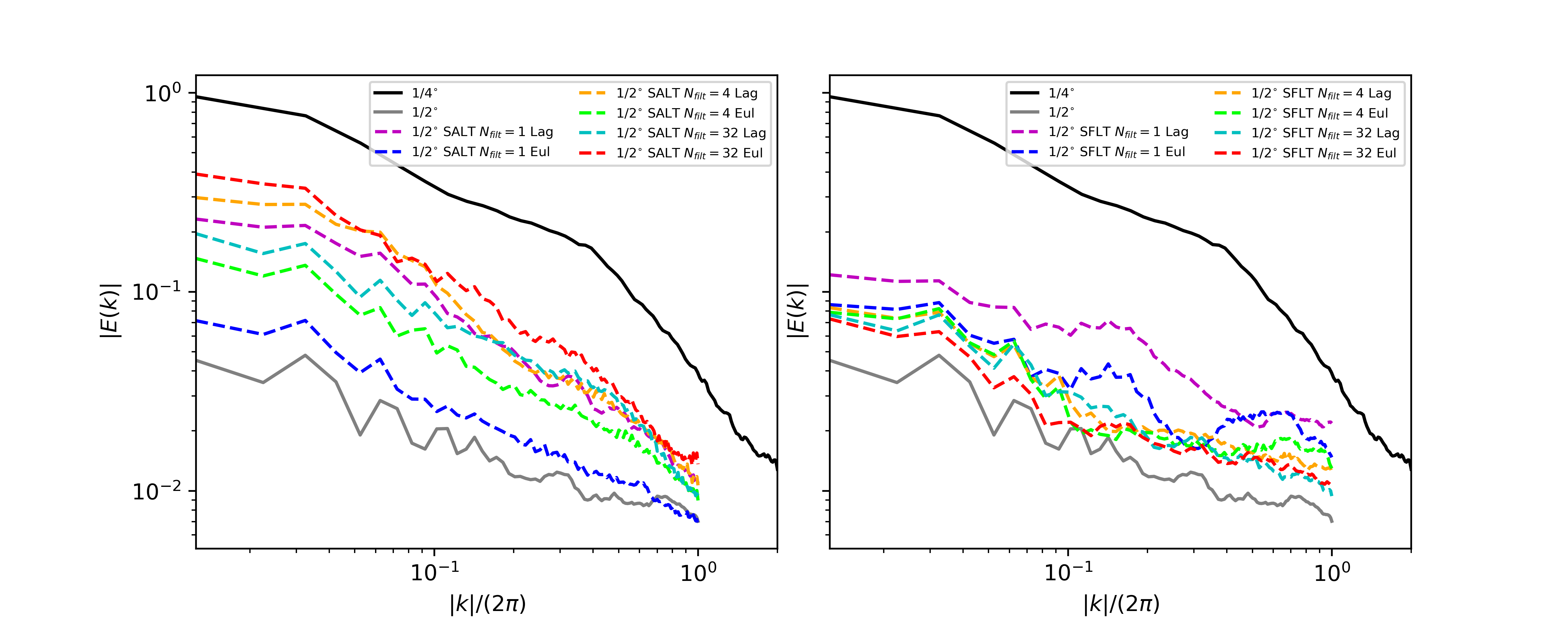}
    \captionsetup{width=0.9\linewidth}
    \caption{Spectra of eddy kinetic energy for SALT (left panel) with $\bs{\xi}_i$ calculated from Lagrangian paths and from Eulerian differences; and for SFLT (right panel) with $\bs{\phi}_I$ calculated from Lagrangian paths and from Eulerian differences. Also included in each plot are the spectra for the deterministic runs on the fine and coarse grids. Spectra are calculated in the $x$-direction at fixed $y=45\frac{1}{6}^\circ$ by $\abs{\hat{E}(k)} := \frac{1}{t_{max}}\int_0^{t_{max}} \abs{\int E(x,t) e^{ikx} dx} \rmd t$. Here $t_{max}=10$years and $t=0$ corresponds to the beginning of the model run, after spin-up. }  %$\abs{\hat{E}(\B{k})} := \frac{1}{T}\int_0^T \abs{\int E(\B{x},t) e^{i\B{k}\cdot\B{x}} d^2x} dt$. }
    \label{fig:SALT_SFLT_spectra}
\end{figure}
\begin{figure}
    \centering
    \includegraphics[width=1.0\textwidth]{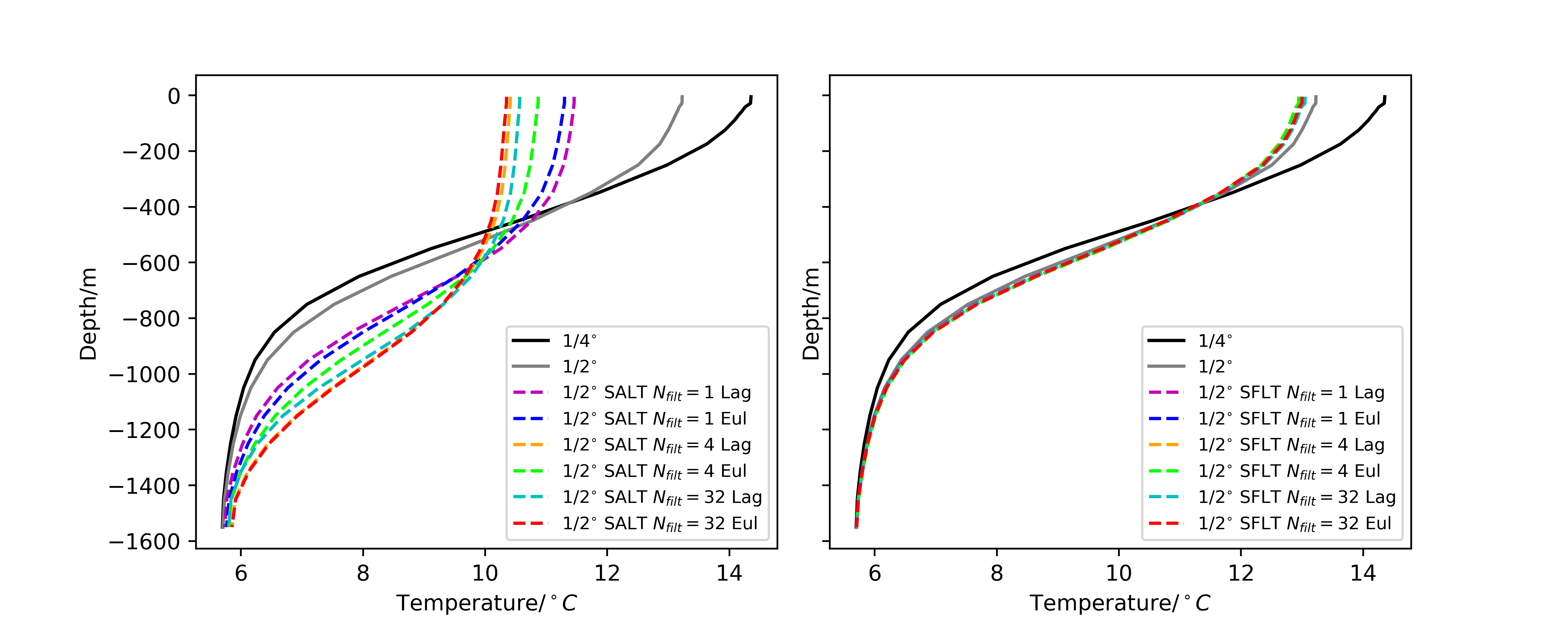}
    \captionsetup{width=0.9\linewidth}
    \caption{Vertical profiles of temperature horizontally-averaged across the domain after 10 years of model time. The left-hand panel shows the results from the SALT runs, alongside the deterministic runs. The right-hand panel shows the results from the SFLT runs, alongside the deterministic runs.}
    \label{fig:temp_SALT_SFLT}
\end{figure}

The results of these runs are shown in \cref{fig:eke_SALT_SFLT,fig:SALT_SFLT_spectra,fig:temp_SALT_SFLT}. \Cref{fig:eke_SALT_SFLT} shows the eddy kinetic energy (EKE), defined by $E = \half \abs{\B{u} - \EXP{\B{u}}}^2$, where $\EXP{\B{u}}$ is the time-averaged velocity. We notice that the eddy kinetic energy is significantly less in the coarse-grid deterministic run than it is in the fine-grid run. This is probably due to the fact that small scales are less present in the coarse-grid flow, and in the coarse-grid model the viscosity used is greater and so kinetic energy is dissipated at a faster rate. However, when we include SALT there is, for most choices of $\bs{\xi}_i$, a notable increase in EKE across the domain, particularly around the western boundary. The exception is in the cases in which the coarse-grained velocities $\bar{\B{v}}$ used to calculate $\bs{\xi}_i$ are defined with only one application of the smoothing operator, as shown in panels (c) and (d) in \cref{fig:eke_SALT_SFLT}. This could be because, from \cref{fig:spectra_xi}, the inclusion of 32 $\bs{\xi}_i$ captures a smaller amount of the total variability; it may also be that the effect of the $\xi_i$s is more spread out across the domain, as shown in \cref{fig:xis}, which overall has a greater impact than having them more highly concentrated in one region. For SFLT there is only a modest improvement in the EKE field, and the effect is similar for all choices of the parameters. In all cases there appears to be little difference between the Eulerian differences method and the Lagrangian paths method when the same $N_{filt}$ is used.
\par
We can also consider the spatial spectra, as shown in \cref{fig:SALT_SFLT_spectra}. There we see that the $1/4^\circ$ run contains higher EKE at all scales than the low-resolution run. Every SALT run succeeds in increasing the energy at almost all scales and in shifting the spectrum towards that of the $1/4^\circ$ run. The most significant improvements are seen in the run with Eulerian parameters computed with $N_{filt}=32$; in contrast, there is only a small change from the deterministic run when the $N_{filt}=1$ Eulerian parameters are used. For SFLT the improvement is again less noticeable, with all choices of parameters only giving a slight increase in EKE at all scales.\\
\par
\begin{figure}
    \centering
    \includegraphics[width=1.0\textwidth]{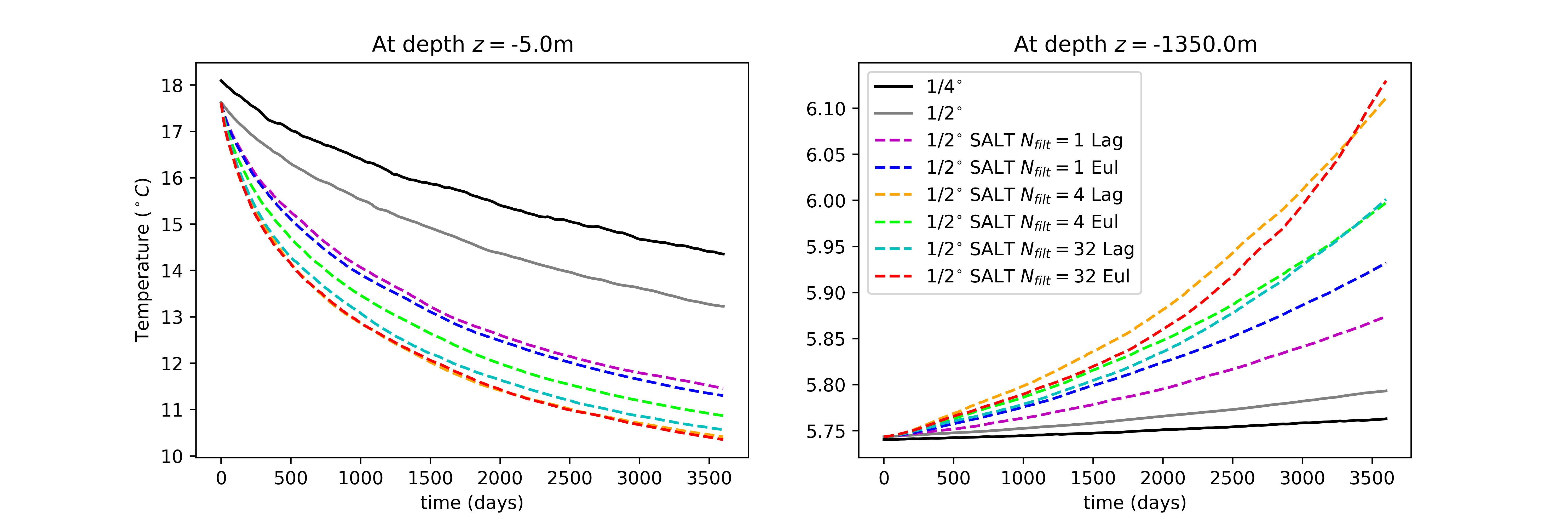}
    \captionsetup{width=0.9\linewidth}
    \caption{Time series of spatially-averaged temperature fields for SALT runs at $z=-5m$ (left panel) and $z=-1350m$ (right panel). }
    \label{fig:temp_time_series}
\end{figure}

Since we are working with the Primitive Equations, the buoyancy can have a large effect on the fluid flow. We therefore consider the temperature, which determines buoyancy directly via the linear equation of state. \Cref{fig:temp_SALT_SFLT} shows vertical temperature profiles at the end of the ten-year run. In the coarse-grid model there is a slightly lower average temperature in the upper layers of the fluid, and slightly higher temperatures in the lower layers. However, with SALT included there is, for some choices of parameters, a significant reduction in temperature in the upper layers, while at lower depths the temperature increases relative to the deterministic model. Considering the time series of spatially averaged temperature at $z = -5m$ and $z=-1350m$ in \Cref{fig:temp_time_series}, we see the downwards diffusion effects are persistent in time. In the deterministic case we see that the coarse-grid model has a stronger downwards diffusion of temperature than the fine-grid run. The inclusion of SALT also accelerates this downward-diffusion effect.
It therefore appears that the calibrated stochastic terms we have included in the temperature equation with SALT cause a downwards-diffusion effect. Indeed, an additional SALT run (not shown), in which the stochastic terms were not included in the temperature advection, did not display this downwards diffusion behaviour. Thus, further investigation will be required in order to determine how to avoid the excessive downwards diffusion in the tracer equation while maintaining a positive effect on the EKE field. SFLT has very little effect on the temperature field when compared to that of the low-resolution model. This is expected however since there are no direct stochastic effects in the temperature equation. Comparing with SALT runs where the temperature downwards-diffusion effect is present against the SFLT runs, we believe that the temperature is the dominant force for the evolution of velocity, at least at the resolutions we have considered here. Then, the limited effects on EKE by the SFLT framework are explained as it does not affect the driving temperature fields directly. It remains part of future work to consider the case where SFLT is added to the temperature field.

\section{Summary and Discussion}
This work lays the groundwork for the application of two relatively new stochastic parameterisation frameworks to the Primitive Equations. The first, SALT, has hitherto only been applied to simple idealised ocean models such as QG and 2D Euler. The second, SFLT, had not been investigated numerically prior to the present work. We have demonstrated some of the desirable theoretical properties of the stochastic Primitive Equations with the noise added in these ways. Notably, the preservation of a circulation theorem for SALT and energy conservation for SFLT. We have proposed to calculate the parameters $\bs{\xi}_i$ governing SALT and $\bs{\phi}_I$ governing SFLT by two different methods: Lagrangian paths and Eulerian differences. We find that there are no significant differences between the two methods, either in the parameters themselves or in the results of model runs. In this case it is preferable to use the Eulerian differences method, as the parameters in this case are computationally less expensive to compute. However, we have used a set-up in which the fine-grid resolution is only is only 2 times the coarse-grid resolution. However, using a larger ratio of grid resolutions would mean more time-steps are needed in the Lagrangian paths and so may give different EOFs that differ more significantly than what we have observed here. We do observe, however, that there are large sensitivities to the choice of smoothing used in defining the coarse-grained velocity, from which the parameters are calculated. In the SALT case, the model runs using parameters calculated with a strong smoothing filter show a significant improvement in the eddy kinetic energy field at all depths, as well as in the eddy kinetic energy spectrum. In the SFLT case, the improvement in EKE field and EKE spectrum are more modest compared to the improvement by SALT due to the lack of direct stochastic effects to the driving temperature fields. Considering the temperature profile, however, we observe that SALT causes significant additional downward diffusion when compared with the deterministic model. It remains an open problem to devise a method to avoid this effect. The answer may lie in a different method for configuring the parameters $\bs{\xi}_i$ or it may be the case that this is a property intrinsic to SALT. In either case, further study is needed in this direction.
\par
% It will also be necessary to determine in future work whether the effects observed in these stochastic parameterisations are dependent on the mesh resolution. As resolution increases we expect the influence of viscosity to diminish and so there may be further effects from SALT and SFLT which in low-resolution simulations are dissipated by the viscosity. 
The stochastic parameterisation frameworks considered in this paper distils all uncertainties of the ocean models into the stochastic parameters $\bs{\xi}_i$ and $\bs{\phi}_I$.
However, the effects of these stochastic parameterisations could be limited by the model, both physically and numerically. Examples of the limiting factors for the Primitive Equations are the forcing from the temperature field and artificial viscosity imposed for numerical stability. The interplay between numerical effects such as artificial viscosity and stochastic parameterisation is particularly interesting for future work. This is due to different numerical viscosity are imposed at different mesh resolutions to numerical stability which influences the calibration process.
%.{\color{red} SP: What do you mean by imposing numerical viscocity before and after the calibration? \\RH: the calibration is done with the fine grid results which is computed with a smaller viscosity than the coarse grid runs.} 
Thus, we expect there are limits to the effects of SALT and SFLT for low-resolution simulations where viscosity are dominant. In high-resolution simulations, we expect to see further effects of stochasticity as the influence of viscosity diminishes. After all, the problem of stochastic parameterisations are not just model-dependent, it also dependent on the numerical method solving it.

\subsection*{Acknowledgements} 
We are grateful to our friends and colleagues who have generously offered their time, thoughts, and encouragement in the course of this work during the time of COVID-19. Thanks to P. Berloff, C. Cotter, S. Danilov, D. Holm, S. Juricke, E. Luesink, W. Pan and all members of the Geometric Mechanics group at Imperial College for their thoughtful comments and discussions. We acknowledge the Alfred Wegener Institute for the use of their computing facilities. The authors are grateful for partial support, as follows. 
% DH for European Research Council (ERC) Synergy grant STUOD - DLV-856408;
RH for the EPSRC scholarship (Grant No. EP/R513052/1); and 
SP for the EPSRC Centre for Doctoral Training in the Mathematics of Planet Earth (Grant No. EP/L016613/1). Finally, we thank the anonymous reviewer for giving useful feedback on our manuscript. 

\appendix

\section{Numerical Implementation}
\label{sec:Numerics}

In order to apply SALT and SFLT to FESOM2 we adapt the time-stepping scheme to include the appropriate stochastic terms. Details of the original (deterministic) time-stepping are given in \citep{danilov_sidorenko_wang_jung_2016}. We modify the scheme from FESOM2 to a two-step Heun-type method \citep{burrage2004}; we choose this because of the use of Stratonovich integrals, to which the Heun method converges. The first step in the method is to compute the modified pressure:
\algn{\begin{split}
            \hat{p}^{n}_h = \rho_0 g \int_{-H}^{z} b(T^{n+1/2}) dz' &+ \sum_i \int_{-H}^z \PD{\xi_i}{z}\cdot\B{u}^n dz'\frac{\Delta W_{n+1}^i}{\Delta t} \\
            &+ \sum_I \int_{-H}^z \PD{\bs{\phi}_I}{z}\cdot\B{u}^n dz'\frac{\Delta B_{n+1}^I}{\Delta t} \end{split}
}
where $\Delta W_{n+1}^i$ and $\Delta B_{n+1}^I$ are independent, normally-distributed random variables with mean $0$ and variance $\Delta t$. For the sake of conciseness we shall assume that the buoyancy depends only on temperature $T$, and that salinity is kept constant; however, extending the method to include additional tracers should be straightforward. The advective, diffusive and pressure parts of the momentum right-hand-side are then computed:
\algn{
    \Delta\hat{\B{u}}^{n+1}  &= \hat{\B{R}}^{n+1/2} - \nabla\lrb{ \hat{p}^n_h + \eta^n}\Delta t + \B{D}\lrb{\B{u}^n,\Delta \hat{\B{u}}^{n+1}} 
}
where $\hat{\B{R}}^{n+1/2}$ is an Adams-Bashforth interpolation of the advective and Coriolis terms. In fact we have $ \hat{\B{R}}^{n+1/2}  = \lrb{\frac32 + \eps}\hat{\B{R}}^{n} - \lrb{\frac12 + \eps}\hat{\B{R}}^{n-1}$, where $ \hat{\B{R}}^n   =  \B{R}\lrsq{\B{v}^n\Delta t,\B{u}^n} +\sum_i\lrb{ \B{R}\lrsq{\bs{\xi}_i,\B{u}^n} -\nabla^{(h)}\bs{\xi}_i \cdot\B{u}^n }\Delta W_{n+1}^i - \sum_I\B{R}\lrsq{\B{v}^n, \bs{\phi}_I\Delta B_{n+1}^I}$ and $\B{R}\lrsq{\B{v},\B{u}}:=-\nabla\cdot\lrb{\B{v}\B{u}} - \B{f}\times\B{v}$. $\B{D}$ includes the horizontal and vertical diffusion terms, as well as the external wind forcing. 
\par
The change in free surface height $\Delta \hat{\eta}^{n+1}$ is computed implicitly:
\algn{
    \lrb{1 - g\Delta^{2}\nabla\cdot\int_{-H}^{0} \nabla \lrb{\cdot } dz }\Delta\hat{\eta}^{n+1}   & =  - \nabla\cdot\int_{-H}^{0} \nabla\cdot\lrb{\B{u}^{n} + \Delta \hat{\B{u}}^{n+1}} dz \Delta t  
}
Once this has been solved we can finally compute the stepped-forward horizontal velocity:
\algn{
    \hat{\B{u}}^{n+1}& = \B{u}^{n} + \Delta\hat{\B{u}}^{n+1} - g\Delta t \nabla\Delta\hat{\eta} 
}   
Then we solve for the total layer thickness $\bar{h}$, which in the continuous case is the same as the free surface height $\eta$; in the discrete case, however, they are different and we compute:
\algnn{
     \hat{\bar{h}}^{n+3/2} &= \bar{h}^{n+1/2} - \nabla\cdot\int_{-H}^{0} \hat{\B{u}}^{n+1} dz \Delta t  
     }
In our present set-up we then set the free-surface height as a linear interpolation of the total layer heights:
\algn{
     \hat{\eta}^{n+1} & = \theta \hat{\bar{h}}^{n+3/2} + (1-\theta)\hat{\bar{h}}^{n+1/2} 
}
where $\theta\in[0,1]$ is an arbitrary parameter, which we set equal to $1$.\\
\par
Since we have the horizontal velocity we may compute the vertical velocity:
\algn{
    \hat{w}^{n+1} &= -\nabla\cdot\int_{-H}^{z} \hat{\B{u}}^{n+1} dz'
}
The newly-computed three-dimensional velocity, along with the stochastic SALT velocity, is then used to advect the tracer:
\algn{
 \hat{T}^{n+3/2} &= T^{n+1/2}  -  R_{T}\lrsq{T^{n+1/2}, T^{n-1/2}, \hat{\B{v}}^{n+1}\Delta t+\xi_i\Delta W_{n+3/2}}   + K\lrsq{T^{n+1/2}}
}
where $R_T$ denotes the advection scheme and $K$ is the diffusion. From these steps we compute intermediate values $\hat{X}^{n+1} := \lrb{\hat{\B{u}}^{n+1}, \hat{\eta}^{n+1}, \hat{\bar{h}}^{n+3/2}, \hat{T}^{3/2}}$ from values at the previous two time steps: $\B{u}^n, \B{u}^{n-1}, \bar{h}^{n+1/2}, T^{n+1/2},T^{n-1/2}$. We may write this schematically as:
\algn{
    \hat{X}^{n+1} = X^n + \mathcal{F}\lrsq{X^n, X^{n-1}}
}
where $\mathcal{F}$ is an operator representing the computations outlined above. For the corrector step we follow the same steps as above, to compute $\mathcal{F}\lrsq{\hat{X}^{n+1}, X^n}$ and we have the overall evolution given by:
\algn{
         X^{n+1} = X^{n} + \Half\lrsq{ \mathcal{F}\lrsq{X^n, X^{n-1}} + \mathcal{F}\lrsq{\hat{X}^{n+1}, X^n} } 
}
This method differs from the usual Heun method because the right-hand side depends on the previous two time-steps, rather than just the previous one. It remains to prove that adding the stochasticity with this method does converge to the required Stratonovich integrals.

\newpage
  \bibliographystyle{alpha} 
  \bibliography{refs}

\newcommand{\etalchar}[1]{$^{#1}$}
\begin{thebibliography}{MAH{\etalchar{+}}97}

\bibitem[BBT04]{burrage2004}
K.~Burrage, P.~M. Burrage, and T.~Tian.
\newblock Numerical methods for strong solutions of stochastic differential
  equations: an overview.
\newblock {\em Proceedings of the Royal Society of London. Series A:
  Mathematical, Physical and Engineering Sciences}, 460(2041):373--402, 2004.

\bibitem[Ber05]{berloff_2005}
Pavel~S. Berloff.
\newblock Random-forcing model of the mesoscale oceanic eddies.
\newblock {\em Journal of Fluid Mechanics}, 529:71–95, 2005.

\bibitem[CCH{\etalchar{+}}19]{cotter2019numerically}
Colin Cotter, Dan Crisan, Darryl~D. Holm, Wei Pan, and Igor Shevchenko.
\newblock {Numerically Modeling Stochastic Lie Transport in Fluid Dynamics}.
\newblock {\em Multiscale Modeling \& Simulation}, 17(1):192--232, 2019.

\bibitem[CCH{\etalchar{+}}20]{cotter2018modelling}
Colin Cotter, Dan Crisan, Darryl~D. Holm, Wei Pan, and Igor Shevchenko.
\newblock {Modelling uncertainty using stochastic transport noise in a 2-layer
  quasi-geostrophic model}, 2020.

\bibitem[CGH17]{CGH17}
C.~J. Cotter, G.~A. Gottwald, and D.~D. Holm.
\newblock {Stochastic partial differential fluid equations as a diffusive limit
  of deterministic Lagrangian multi-time dynamics}.
\newblock {\em Proceedings of the Royal Society A: Mathematical, Physical and
  Engineering Sciences}, 473(2205):20170388, 2017.

\bibitem[CH09]{Cotter_holm_2009}
C.~J. Cotter and D.~D. Holm.
\newblock {Continuous and Discrete Clebsch Variational Principles}.
\newblock {\em Foundations of Computational Mathematics}, 9(2):221--242, 2009.

\bibitem[{Cot}20]{cotter2020data}
{Cotter, Colin and Crisan, Dan and Holm, Darryl and Pan, Wei and Shevchenko,
  Igor}.
\newblock {Data Assimilation for a Quasi-Geostrophic Model with
  Circulation-Preserving Stochastic Transport Noise}.
\newblock {\em Journal of Statistical Physics}, 179(5):1186--1221, 2020.

\bibitem[Doo53]{doob1953stochastic}
Joseph~Leo Doob.
\newblock {\em Stochastic processes}, volume~10.
\newblock New York Wiley, 1953.

\bibitem[DSWJ16]{danilov_sidorenko_wang_jung_2016}
S.~Danilov, D.~Sidorenko, Q.~Wang, and T.~Jung.
\newblock {The Finite-volumE Sea ice-Ocean Model} ({FESOM2}).
\newblock {\em Geoscientific Model Development Discussions}, pages 1--44, 2016.

\bibitem[GBB{\etalchar{+}}00]{GRIFFIES2000123}
Stephen~M. Griffies, Claus B\"{o}ning, Frank~O. Bryan, Eric~P. Chassignet,
  R\"{u}diger Gerdes, Hiroyasu Hasumi, Anthony Hirst, Anne-Marie Treguier, and
  David Webb.
\newblock {Developments in ocean climate modelling}.
\newblock {\em {Ocean Modelling}}, 2(3):123--192, 2000.

\bibitem[Hal13]{hallberg_2013}
Robert Hallberg.
\newblock {Using a resolution function to regulate parameterizations of oceanic
  mesoscale eddy effects}.
\newblock {\em Ocean Modelling}, 72:92--103, 2013.

\bibitem[HH21]{HolmHu2021}
Darryl~D. Holm and Ruiao Hu.
\newblock Stochastic effects of waves on currents in the ocean mixed layer.
\newblock {\em Journal of Mathematical Physics}, 62(7):073102, 2021.

\bibitem[HJS07]{Hannachi_2007}
A.~Hannachi, I.~T. Jolliffe, and D.~B. Stephenson.
\newblock {Empirical orthogonal functions and related techniques in atmospheric
  science: A review}.
\newblock {\em International Journal of Climatology}, 27(9):1119--1152, 2007.

\bibitem[HLB96]{holmes_lumley_berkooz_1996}
Philip Holmes, John~L. Lumley, and Gal Berkooz.
\newblock {\em Turbulence, Coherent Structures, Dynamical Systems and
  Symmetry}.
\newblock Cambridge Monographs on Mechanics. Cambridge University Press, 1996.

\bibitem[HMR98]{HMR1998}
Darryl~D Holm, Jerrold~E Marsden, and Tudor~S Ratiu.
\newblock {The Euler-Poincar\'{e} Equations and Semidirect Products with
  Applications to Continuum Theories}.
\newblock {\em Advances in Mathematics}, 137(1):1--81, 1998.

\bibitem[Hol15]{holm2015}
Darryl~D. Holm.
\newblock {Variational principles for stochastic fluid dynamics}.
\newblock {\em Proceedings of the Royal Society A: Mathematical, Physical and
  Engineering Sciences}, 471(2176):20140963, 2015.

\bibitem[HSS09]{holm2009geometric}
Darryl~D Holm, Tanya Schmah, and Cristina Stoica.
\newblock {\em Geometric mechanics and symmetry: from finite to infinite
  dimensions}, volume~12.
\newblock Oxford University Press, 2009.

\bibitem[Kor17]{KORN2017525}
Peter Korn.
\newblock {Formulation of an unstructured grid model for global ocean
  dynamics}.
\newblock {\em {Journal of Computational Physics}}, 339:525--552, 2017.

\bibitem[MAH{\etalchar{+}}97]{marshall_adcroft_hill_perelman_heisey_1997}
John Marshall, Alistair Adcroft, Chris Hill, Lev Perelman, and Curt Heisey.
\newblock {A finite-volume, incompressible Navier Stokes model for studies of
  the ocean on parallel computers}.
\newblock {\em Journal of Geophysical Research: Oceans}, 102(C3):5753--5766,
  1997.

\bibitem[Mem14]{memin2014}
Etienne Memin.
\newblock Fluid flow dynamics under location uncertainty.
\newblock {\em Geophysical \& Astrophysical Fluid Dynamics}, 108(2):119--146,
  2014.

\bibitem[Mey62]{meyer1962decomposition}
Paul-Andr{\'e} Meyer.
\newblock A decomposition theorem for supermartingales.
\newblock {\em Illinois Journal of Mathematics}, 6(2):193--205, 1962.

\bibitem[Mey63]{meyer1963decomposition}
Paul-Andr{\'e} Meyer.
\newblock Decomposition of supermartingales: the uniqueness theorem.
\newblock {\em Illinois Journal of Mathematics}, 7(1):1--17, 1963.

\bibitem[PZ14]{mana2014toward}
PierGianLuca {Porta Mana} and Laure Zanna.
\newblock {Toward a stochastic parameterization of ocean mesoscale eddies}.
\newblock {\em Ocean Modelling}, 79:1--20, 2014.

\bibitem[RDH{\etalchar{+}}12]{SOMA}
T~Ringler, D~Danilov, R~Hallberg, A~Adcroft, P~Berloff, and P~Gent.
\newblock {A test case for the assessment of Simulating Ocean Mesoscale
  Activity (SOMA)}.
\newblock {\em Unpublished manuscript.}, 2012.

\bibitem[SC21]{street2021}
O.~D. Street and D.~Crisan.
\newblock Semi-martingale driven variational principles.
\newblock {\em Proceedings of the Royal Society A: Mathematical, Physical and
  Engineering Sciences}, 477(2247):20200957, 2021.

\bibitem[SDB{\etalchar{+}}16]{sein2016designing}
Dmitry~V Sein, Sergey Danilov, Arne Biastoch, Jonathan~V Durgadoo, Dmitry
  Sidorenko, Sven Harig, and Qiang Wang.
\newblock Designing variable ocean model resolution based on the observed ocean
  variability.
\newblock {\em Journal of Advances in Modeling Earth Systems}, 8(2):904--916,
  2016.

\bibitem[SW68]{Seliger1968}
R.~L. Seliger and Gerald~Beresford Whitham.
\newblock Variational principles in continuum mechanics.
\newblock {\em Proceedings of the Royal Society of London. Series A.
  Mathematical and Physical Sciences}, 305(1480):1--25, 1968.

\bibitem[UJPD21]{uchida2021ensemble}
Takaya Uchida, Quentin Jamet, Andrew Poje, and William~K Dewar.
\newblock {An ensemble-based eddy and spectral analysis, with application to
  the Gulf Stream}.
\newblock {\em Journal of Advances in Modeling Earth Systems}, page
  e2021MS002692, 2021.

\end{thebibliography}

\end{document}